\documentclass[preprint]{elsarticle}

\usepackage{lineno,hyperref}
\usepackage{graphicx}
\usepackage{caption}
\usepackage{subcaption}
\usepackage{color}
\usepackage{amsmath}

\modulolinenumbers[200]

\journal{arXiv, to be submitted to AEU INT J ELECTRON C}

\begin{document}

\begin{frontmatter}

\title{On the Security of a Revised Fragile Watermarking Scheme}

\author{Daniel Caragata}
\address{Universidad Tecnica Federico Santa Maria, Departamento de Electronica}
\address{daniel.caragata@usm.cl}

\begin{abstract}
This paper analyzes a revised fragile watermarking scheme proposed by Botta et al. which was developed as a revision of the watermarking scheme previously proposed by Rawat et al. A new attack is presented that  allows an attacker to apply a valid watermark on tampered images, therefore circumventing the protection that the watermarking scheme under study was supposed to offer. Furthermore, the presented attack has very low computational and memory requirements.
\end{abstract}
\end{frontmatter}

\linenumbers

\section{Introduction}
\label{sec:introduction}

In recent years we have witnessed an impressive increase in the quantity of digital images being taken and transmitted, mostly due to the development of affordable digital cameras and to the high penetration rate of the Internet. This context has lead to an augmented effort of the scientific community to develop tools that protect digital images. One of these tools is fragile watermarking, which is supposed to protect the integrity and authenticity of digital images: a watermark is inserted in the image such that there are no perceptible changes to the human eye, but that it is possible to detect any malicious changes to the image. 

Various watermarking schemes have been proposed that work either in the pixel domain \cite{mon1, che1, raw1} or in the transform domain \cite{ali1, xin1, wan1, al1}. As it is normal in cryptography, a lot of effort has also been put in attacking some of the proposed watermarking schemes \cite{bot1, car1, ten1}. 

Rawat et al. have proposed a new fragile watermarking, \cite{raw1}, that embeds the watermark in the LSB of every pixel. However, it was soon attacked by Teng et al.  \cite{ten1} as well as by Botta et al. \cite{bot1} and both papers have also presented revised versions of the algorithm. Teng et al.'s proposed fragile watermarking embeds a watermark, that is also a function of the pixel content, in one of the three least significant bits. A chaotic function is used to determine the exact location of the watermark. Botta et al.'s fragile watermarking scheme generates a watermark that is a function of the pixel value, but always embeds it in the least significant bit of the pixel. This paper studies the security of Botta et al.'s fragile watermarking algorithm and shows that it is relatively easy to break.

The rest of this paper is organized as follows. In Section ~\ref{sec:cryptographic} we present the cryptographic perspective that we will use for the cryptanalysis as well as the different attack models, in order to better understand the cryptographic context of the attack we propose. Section ~\ref{sec:revised} presents the details of Botta et al.'s fragile watermarking and an example of its use. In Section ~\ref{sec:cryptanalysis} we describe the proposed attack and we present one successful implementation of the attack. Section ~\ref{sec:conclusions} concludes our work.

\section{Cryptographic security}
\label{sec:cryptographic}

Fragile watermarking is a cryptographic primitive because it tries to protect the integrity of some data, i.e. the image, in an adversary setting \cite{sch1}. Therefore, in order for a fragile watermarking algorithm to be considered secure, it must respect Kerckhoffs Principle \cite{ker1, paa1}. This means that the algorithm  must be secure in the worst case scenario: when the attacker knows all the details of the algorithm except the secret key that was being used. Also, the algorithm must be secure in all of the 4 most common models of attack \cite{sch1}:
\begin{itemize}
\item \textbf{ciphertext only model}: the attacker only has access to some \textit{protected data}.
\item \textbf{known plaintext model}: the attacker  has access to some \textit{protected data} as well as the corresponding \textit{unprotected data}.
\item \textbf{chosen plaintext model}: the attacker is able to obtain the corresponding \textit{protected data} of some \textit{(unprotected) data} he chooses.
\item \textbf{chosen ciphertext model}: the attacker is able to submit some \textit{protected data} to the cryptographic primitive under attack and can obtain the corresponding \textit{unprotected data} or the verification of its integrity.
\end{itemize}

In the above enumeration we have used the terms \textit{"protected data"} and \textit{"unprotected data"} in order to make the definitions as broad as possible. For example, if the cryptographic primitive under study is an encryption algorithm, the \textit{"protected data"} is the ciphertext, while the \textit{"unprotected data"} is the plaintext. 

This paper will present a chosen plaintext attack on a recently proposed fragile watermarking algorithm. In the case of fragile watermarking, the term \textit{"protected data"} refers to the watermarked image, while the term \textit{"unprotected data"} refers to the unprotected image, i.e. the image prior to the embedding of the watermark.

\section{Revised fragile watermarking algorithm}
\label{sec:revised}

Botta et al. have recently proposed a fragile watermarking algorithm \cite{bot1} based on the cryptanalysis of Rawat et al's fragile watermarking algorithm \cite{raw1}. The proposed fragile watermarking algorithm embeds the watermark as follows:
\begin{enumerate}[E1.]
\item The original image, $I_{h}$, of size $m \times n$ is scrambled using Arnold cat map \cite{arn1} \textit{k} times, to obtain the scrambled image $I_{s}$. 
\item A chaotic sequence, \textit{C}, is generated using logistic map \cite{may1}. The sequence is of same size as the image $I_{h}$. Further, the values of \textit{C} are rounded off to obtain a bit sequence.
\item For every pixel $p_{i}$, of coordinates $x$ and $y$, the binary chaotic watermark $W_{c}(x,y)$ is computed as:
\begin{equation} \label{eq:1}
W_{c}(x,y)=parity(W(x,y)\oplus C(x,y)\mid MAC_{k}(p_{i})AND  0 \times FE \mid x \mid y))
\end{equation}
where "\textit{parity}" means the number of bits with value one, \textit{W} represents the original watermark, MAC is a Message Authentication Code and "$\mid$" means concatenation.
\item Embed $W_{c}$ in the LSB (Least Significant Bit) of every pixel of $I_{s}$.
\item To obtain the watermarked image, $I_{w}$, Apply Arnold cat map \textit{T-k} times, where \textit{T} is the period of Arnold cat map.
\end{enumerate}

For watermark extraction from a watermarked image, $I_{w}$, the following steps need to be performed:
\begin{enumerate}[D1.]
\item The watermarked image is scrambled applying Arnold cat map \textit{k} times, to obtain the image $I_{ws}$.
\item The binary chaotic sequence \textit{C} is computed exactly the same way as for the embedding process.
\item For every pixel $p_{i}$, of coordinates $x$ and $y$, an expected watermark is computed as:
\begin{equation} \label{eq:2}
W_{e}(x,y)=parity(W(x,y)\oplus C(x,y)\mid MAC_{k}(p_{i})AND 0 \times FE \mid x \mid y))
\end{equation}
\item A new image, $I_{temp}$, is constructed as the absolute difference between the expected watermark, $W_{e}(x,y)$ and the LSB of every pixel of $I_{ws}$.
\item The tampered regions are detected applying Arnold cat map \textit{T-k} times on $I_{temp}$.
\end{enumerate}

\begin{figure}
\centering

\begin{subfigure}{.5\textwidth}
  \centering
  \includegraphics[width=.4\linewidth]{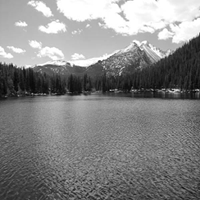}
  \caption{Original image}
  \label{fig:sub1a}
\end{subfigure}

\begin{subfigure}{.5\textwidth}
  \centering
  \includegraphics[width=.4\linewidth]{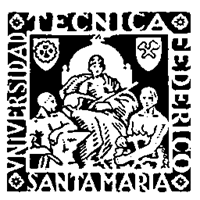}
  \caption{Image watermark}
  \label{fig:sub2a}
\end{subfigure}

\begin{subfigure}{.5\textwidth}
  \centering
  \includegraphics[width=.4\linewidth]{nomoim}
  \caption{Watermarked image}
  \label{fig:sub3a}
\end{subfigure}

\begin{subfigure}{.5\textwidth}
  \centering
  \includegraphics[width=.4\linewidth]{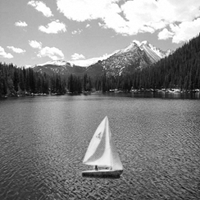}
  \caption{Tampered image}
  \label{fig:sub4a}
\end{subfigure}

%\begin{subfigure}{.5\textwidth}
%  \centering
%  \includegraphics[width=.4\linewidth]{exwaim}
%  \caption{Extracted watermark}
%  \label{fig:sub5a}
%\end{subfigure}

\begin{subfigure}{.5\textwidth}
  \centering
  \includegraphics[width=.4\linewidth]{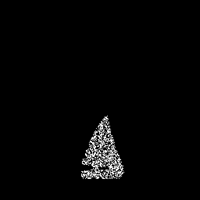}
  \caption{Detected tampered images}
  \label{fig:sub6a}
\end{subfigure}

\caption{Botta et al's fragile watermarking algorithm}
\label{fig:figwm}
\end{figure}

We have implemented Botta et al's watermarking algorithm and we present the results in Figure ~\ref{fig:figwm}.

Figure ~\ref{fig:sub1a} shows the original image, which has been watermarked with the watermark image shown in Figure ~\ref{fig:sub2a}. The resulting watermarked image is presented in Figure ~\ref{fig:sub3a}. We have tampered the watermarked image by inserting a boat on the lake, as shown in Figure ~\ref{fig:sub4a}. Then, we have applied the watermark detection and we have obtained the tampered regions presented in Figure ~\ref{fig:sub6a}. We can conclude that the watermarking algorithm performs as expected.

\section{Cryptanalysis}
\label{sec:cryptanalysis}
In this section we will present an attack against Botta et al's fragile watermarking scheme. We will first describe the attack and then show the results of its implementation

\subsection{Attack description}
\label{subsec:attack}
The attack presented in this paper falls under the chosen plaintext model, i.e. the attacker needs to be able to obtain the watermarked version of some images he chooses.

The watermarking algorithm under study is supposed to be secure because the embedded watermark in every pixel is a function of the secret key, the 7 most significant bits of the pixel and of the pixel coordinates. However, this means that if a pixel in a given location in the image, $(x, y)$, has the same value in different images, the same watermarking bit will be embedded. Moreover, since the LSB of the pixel is used to embed the watermark, only $2^{7}=128$ possible values exist for every pixel position. 

Therefore, the attacker needs to know the watermarking bit $W_{c}(i)$ for every pixel $p_{i}$, of coordinates $x$ and $y$. For this, he will run the watermarking algorithm 128 times and will store the LSB plane of the watermarked image. With this information, the attacker will construct a three dimensional look-up table whose search fields will be the two coordinates, $(x, y)$, and the value of the 7 most significant bits of the pixel, and whose values will be the corresponding watermarking bit $W_{c}(i)$. This look-up table requires $m \times n \times 128$ bits storage space.

Once the attacker has constructed the above mentioned look-up table, he can apply a valid watermark on any image. Therefore, the attacker has the ability to circumvent the protection offered by the watermarking algorithm under study.

\subsection{Implementation results}
\label{subsec:implementation}

\begin{figure}
\centering
\begin{subfigure}{.5\textwidth}
  \centering
  \includegraphics[width=.4\linewidth]{nomoim}
  \caption{Watermarked image}
  \label{fig:sub1}
\end{subfigure}
\begin{subfigure}{.5\textwidth}
  \centering
  \includegraphics[width=.4\linewidth]{waim}
  \caption{Image watermark}
  \label{fig:sub2}
\end{subfigure}
\begin{subfigure}{.5\textwidth}
  \centering
  \includegraphics[width=.4\linewidth]{moim}
  \caption{Attacked tampered image}
  \label{fig:sub3}
\end{subfigure}

%\begin{subfigure}{.5\textwidth}
%  \centering
%  \includegraphics[width=.4\linewidth]{waim}
%  \caption{Extracted watermark}
%  \label{fig:sub4}
%\end{subfigure}

\begin{subfigure}{.5\textwidth}
  \centering
  \includegraphics[width=.4\linewidth]{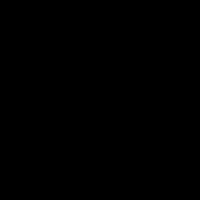}
  \caption{Detected tampered regions}
  \label{fig:sub5}
\end{subfigure}

\caption{Unsuccessful detection of tampered image}
\label{fig:figattack}
\end{figure}

We have implemented the proposed attack and we were able to apply a valid watermark on a tampered image, as shown in Figure ~\ref{fig:figattack}. 
 
Figure ~\ref{fig:sub1} shows the image watermarked with the watermark image presented in Figure ~\ref{fig:sub2}. We have tampered this image by inserting a boat in the lake and we have used the attack presented in the previous subsection to apply a valid watermark on the tampered image. The resulting image is shown in ~\ref{fig:sub3}. When running the watermark detection algorithm, the LSB of all the pixels will be equal to the expected watermark computed using Eq. \eqref{eq:2}, therefore the watermarking algorithm will not detect that there are tampered regions, as shown in Figure ~\ref{fig:sub5}.

\section{Conclusions}
\label{sec:conclusions}

This paper has analyzed the security of a recently proposed fragile watermarking scheme and has presented an attack that is capable to circumvent the protection that the watermarking scheme under study is supposed to offer. Furthermore, the computational complexity of the attack is very small, only 128 runs of the watermarking scheme are needed. Also, the memory requirements are very limited, only  $m \times n \times 128$ bits are needed. We have abstained ourselves from proposing an improvement to the scheme because we feel there are a lot of available watermarking schemes and that the scientific community would benefit more from a thorough analysis of the already proposed watermarking schemes.

\section{Acknowledgments}
\label{sec:acknowledgments}

Financing from DGIP-UTFSM 23.15.37 research project is gratefully acknowledged.

\section*{References}
\label{sec:references}

\bibliography{paperbib}

\begin{thebibliography}{10}

\bibitem{mon1}
M.~Moniruzzaman, M.~Hawlader, and M.~Hossain, ``An image fragile watermarking
  scheme based on chaotic system for image tamper detection,'' in {\em 2014
  International conference on Informatics, Electronics and Vision, ICIEV 2014},
  (Dhaka, Bangladesh), 2014.

\bibitem{che1}
F.~Chen, H.~He, H.~Tai, and H.~Wang, ``Chaos-based self-embedding fragile
  watermarking with flexible watermark payload,'' {\em Multimedia Tools and
  Applications}, vol.~72, pp.~41--56, 2014.

\bibitem{raw1}
S.~Rawat and B.~Raman, ``A chaotic system based fragile watermarking scheme for
  image tamper detection,'' {\em International Journal of Electronics and
  Communications (AEU)}, vol.~65, pp.~840--847, 2011.

\bibitem{ali1}
M.~Ali, C.~Ahn, M.~Pant, and P.~Siarry, ``An image watermarking scheme in
  wavelet domain with optimized compensation of singular value decomposition
  via artificial bee colony,'' {\em Information Sciences}, vol.~301,
  pp.~44--60, 2015.

\bibitem{xin1}
L.~Xing, P.~Liu, and L.~Jiang, ``A fragile watermarking scheme for temper
  detection,'' {\em Journal of Information and Computational Science},
  vol.~11:4, pp.~1335--1342, 2014.

\bibitem{wan1}
S.~Wang, D.~Zheng, J.~Zhao, W.~Tam, and F.~Speranza, ``Adaptive watermarking
  and tree structure based image quality estimation,'' {\em IEEE Transactions
  on Multimedia}, vol.~16, pp.~311--325, 2014.

\bibitem{al1}
A.~Al-Haj and A.~Amer, ``Secured telemedicine using region/based watermarking
  with tamper localization,'' {\em Journal of Digital Imaging}, vol.~27,
  pp.~737--750, 2014.

\bibitem{bot1}
M.~Botta, D.~Cavagnino, and V.~Pomponiu, ``A successful attack and revision of
  a chaotic system based fragile watermarking scheme for image tamper
  detection,'' {\em International Journal of Electronics and Communications
  (AEU)}, vol.~69, pp.~242--245, 2015.

\bibitem{car1}
D.~Caragata, A.~Radu, S.~El~Assad, and C.~Apostol, ``Chaos based fragile
  watermarking algorithm for jpeg images,'' in {\em International Conference
  for Internet Technology and Secured Transcations}, (London, UK), pp.~1--7,
  2010.

\bibitem{ten1}
L.~Teng, X.~Wang, and W.~X., ``Cryptanalysis and improvement of a chaotic
  system based fragile watermarking scheme,'' {\em International Journal of
  Electronics and Communications (AEU)}, vol.~67, pp.~540--547, 2013.

\bibitem{sch1}
N.~Ferguson, B.~Schneier, and T.~Kohno, {\em Cryptography Engineering: Design
  Principles and Practical Applications: Design Principles and Practical
  Applications}.
\newblock Wiley, 2011.

\bibitem{ker1}
A.~Kerckhoffs, ``La cryptographie militaire,'' {\em Journal des sciences
  militaires}, vol.~IX, pp.~5--83, 161--191, 1883.

\bibitem{paa1}
C.~Paar and J.~Pelzl, {\em Understanding Cryptography: A Textbook for Students
  and Practitioners}.
\newblock Springer, 2010.

\bibitem{arn1}
V.~Arnold and A.~Avez, ``Problèmes ergodiques de la mecanique classique,''
  {\em Monographies internationales de mathematiques modernes}, vol.~9, 1967.

\bibitem{may1}
R.~May, ``Simple mathematical models with very complicated dynamics,'' {\em
  Nature}, vol.~261(5560), pp.~459--467, 1976.

\end{thebibliography}

\end{document}